# FoamFactor: Hydrogel-Foam Composite with Tunable Stiffness and Compressibility

**Humphrey Yang†[1], Zeyu Yan†[2], Danli Luo[1], Lining Yao[1]**

[1]Human-Computer Interaction
Carnegie Mellon University
Pittsburgh, Pennsylvania, U.S.A.
{hanliny, danlil, liningy}@email.com

[2]Computer Science
University of Maryland
College Park, Maryland, U.S.A.
zeyuy@umd.edu

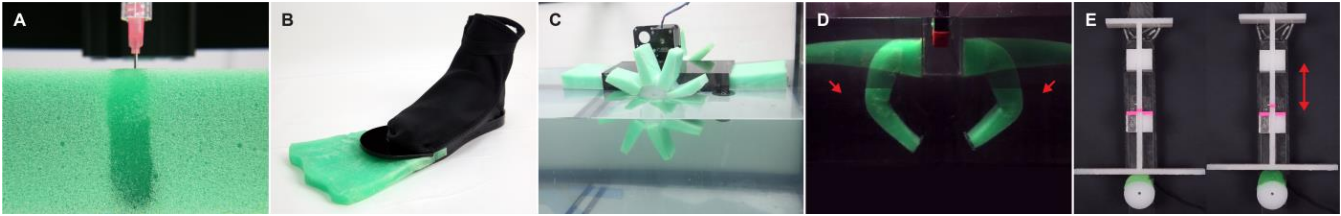

**Figure 1: (A)** Hydrogel injection into a foam matrix to form a hydrogel-foam composite. **(B)** A shoe on land can turn into a fin underwater. **(C)** An amphibious car with round wheels on land can turn into pedals underwater. **(D)** Self-deploying underwater gripper. **(E)** A stiffness-tunable cam wheel that affects the plunger movement.

## ABSTRACT

This paper presents FoamFactor, a novel material with tunable stiffness and compressibility between hydration states, and a tailored pipeline to design and fabricate artifacts consisting of it. This technique compounds hydrogel with open-cell foams via additive manufacturing to produce a water-responsive composite material. Enabled by the large volumetric changes of hydrogel dispersions, the material is soft and compressible when dehydrated and becomes stiffer and rather incompressible when hydrated. Leveraging this material property transition, we explore its design space in various aspects pertaining the transition of hydration states, including multi-functional shoes, amphibious cars, mechanical transmission systems, and self-deploying robotic grippers.

### ACM Classification Keywords
H.5.m. Information interfaces and presentation (e.g., HCI)

## INTRODUCTION

There are two common transmission media for fluid-driven soft actuators - air and water. Compared with pneumatically-driven soft actuators, hydraulic actuators have the advantage of being incompressible and more powerful. In particular, soft hydraulic actuators have been used heavily in underwater soft robots since water is readily available in its context, and water as a transmission fluid enables the actuator to be neutrally buoyant [17]. However, for soft robots on land, it is often challenging to use hydraulic actuators as water is heavy, and a water chamber is needed as water is not available in the context. The breach of the chamber may also impair the robot locomotion. Identically, breaching problems as well as the increased surrounding pressure also causes using pneumatic actuators underwater difficult.

FoamFactor proposes a novel composite material with tunable physical properties to address the previously mentioned issues. Leveraging the properties of its components - sodium polyacrylate (SPA) hydrogel and polyurethane (PU) foam, the material is lightweight, compressible, and conformable when dried, and becomes neutrally buoyant, more incompressible, and stiffer when hydrated. These tunable properties make it a good candidate for an amphibious soft actuator. Situating FoamFactor in HCI literature of material-driven shape changing interfaces, it provides unique features and enriches the existing library of actuation and stiffness-changing materials. Compared with jamming [24], heat [12], electrical or magnetic fields [32] as triggers for stiffness-changing, FoamFactor is untethered and triggered by water. It is unique and efficient for the application contexts that involve water. Additionally, the fabrication process and control methods are straightforward, and the materials are commercially available. Lastly, comparing to other stiffness-tunable materials, the additional tunable compressibility allows for new design spaces (e.g. compact actuators that can expand and self-deploy in water). The major contributions of this paper are explained as follows.

- **A novel hydration-triggered stiffness- and compressibility-changing composite material.** The mechanisms behind the property changes are explained in this paper. Experiments were conducted to characterize the performances and behaviors of this material.
- **A design and digital fabrication workflow adapted to the material.** This workflow adopts commercially available equipment and a CNC gantry system installed with a pneumatic extruder for fabrication, and a customized toolpath-planner for design and G-code generation. The parameters for this workflow were also characterized.
- **Applications to demonstrate the applicability of the material.** We demonstrate this material in various contexts pertaining the transition of hydration states, including multi-functional shoes, amphibious cars, mechanical transmission systems, and self-deploying robotic grippers.

## RELATED WORK

### Stiffness-Changing Interfaces

Jamming materials can switch from fluid-like states to solid-like states by controlling the differential air pressure in the system. Layered sheets (e.g. sand papers) were used to create thin sheet-like materials that can change stiffness [24]. An application that has been explored was wearable haptic feedbacks [30, 31]. Additionally, Follmer S. et al. [10] explored hydraulic jamming for interface design and proposed use cases such as tunable clay, transparent haptic lenses, and ShapePhones. Beyond HCI, pneumatic and hydraulic jamming have also been heavily used in robotics. Most of these jamming bags contained granular jamming medium and had the form of pads [40], tubes [5, 40], or balls [2]. In addition to jamming techniques, perceptual stiffness-changing can be induced by kinetic 2.5D pin displays [19], or by utilizing field-activated materials such as magnetorheological (MR) or electrorheological (ER) fluids [32].

Most of the previous works were tethered to their energy stimuli (e.g. pneumatic or hydraulic control, magnetic field, electrical field, motorized actuation). Compared to them, FoamFactor has a unique stiffness-changing mechanism - hydration - and is untethered. Furthermore, our technique has an advantage of being highly compressible when dried. In this paper, we leveraged these material properties to explore its design space that involves hydration changes.

### Hydration-based Shape-Changing

Natural organisms such as pinecones [26], erodium seeds [9], and wheat awns [7] can transform in response to changes in relative humidity of the atmosphere or upon hydration when entering water. Inspired by how these natural composite materials had differential swelling rates that induced shape-changing, synthetic hygroscopic composite materials were developed to self-transform in water [8]. HCI researchers had also leveraged these mechanisms to design hydration-based shape-changing interfaces. Transformative Appetite [35] investigated differential swelling-induced transformation of foods. BioLogic [38] introduced bacterial actuators that shrink or expand in response to the relative humidity in the environment.

The hydrogel component in FoamFactor is one of such hydration-based shape-changing material. While hydrogel [11, 39] and hydrogel composites [8] had been used for shape changing, the unique combination of foam as matrix and hydrogel as its dispersion of FoamFactor is novel. In its hydrated state, FoamFactor leverages the mechanical advantages of both components to create a stiffer material.

### Foam for Shape-Changing

Foams have been used for shape changing as it is lightweight, flexible, compressible, and can be compounded into composites that are multifunctional. It has been used as the body of a soft robot wrapped by a sensing fabric [4]. Combined with silicone skin layer, foam was used to design pneumatic soft robots [27] and fluid pumps [18]. Some multifunctional foam-based composite materials can self-heal and have changeable stiffnesses [33]. Conductive foams were also explored for sensing applications [20]. In FoamFactor, foams are introduced for a unique purpose. The open-cell foam we chose resists tension well and functions as a matrix material for hydrogel dispersions. The foam is lightweight and highly compressible, ideal for compact packaging scenarios. Foam also provided a stronger and constrained skeleton to hydrated hydrogel, and enabled tendon-driven actuations.

### Material-Driven Shape-Changing Interfaces

Material-driven interfaces that have dynamically variable shapes or other physical properties is an emerging field in HCI. Pneumatic-driven elastomer or non-elastic air bladders have been explored in PneUI [37], Printflatables [29], and aeroMorph [23]. Printed Paper Actuator [34] explored thermoplastic-graphene composite as actuator and paper as substrate for shape-changing. Thermoplastic-based shape-changing materials had also been explored in uniMorph [13], Thermoph [1], and Foldio [22]. PH responsive materials had been explored in Organic Primitives [16]. Liquid metals were introduced for tunable tactile and weight perceptions [21, 28]. Shape memory alloys have been widely explored in HCI works including Surflex [6] and Electronic Popables [25]. 3D printed metamaterials were introduced for tunable textures [15] and compliant mechanisms [14]. FoamFactor is proposed to enrich the existing material library of designing shape-changing interfaces.

## OVERVIEW

### Techniques

We developed a novel hydrogel-foam composite material (FoamFactor) that exhibits two distinct mechanical behaviors when hydrated and dehydrated. To design and fabricate differently structured artifacts with the hydrogel-foam composite, we developed a design tool (Figure 2B) and a customized printing platform (Figure 2C) for fabrication. Figure 2D shows some printed samples. Depending on the hydrogel distribution in the foam, the objects exhibited different deformation behaviors.

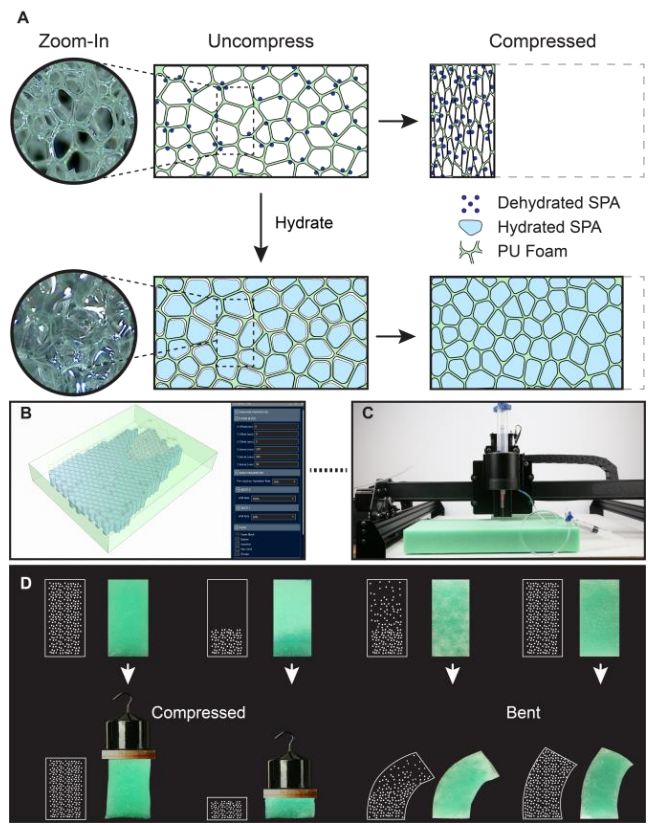

**Figure 2: (A) Mechanisms of the hydrogel-foam composite material. (B) Design tool. (C) Tailored digital fabrication platform. (D) Samples with partial injection showing different compression (left) and bending (right) behaviors, demonstrating stiffness tunability.**

### Design Space

Figure 3 provides an outline of FoamFactor's valuable material behaviors and their potential applications. The properties of the hydrogel-foam composite come from the inherent properties of foam and hydrogel. FoamFactor is designed to reveal the transition of shape and mechanical property in contact with water- the most ubiquitous substance in the world, providing an alternative to water-triggered interaction modality.

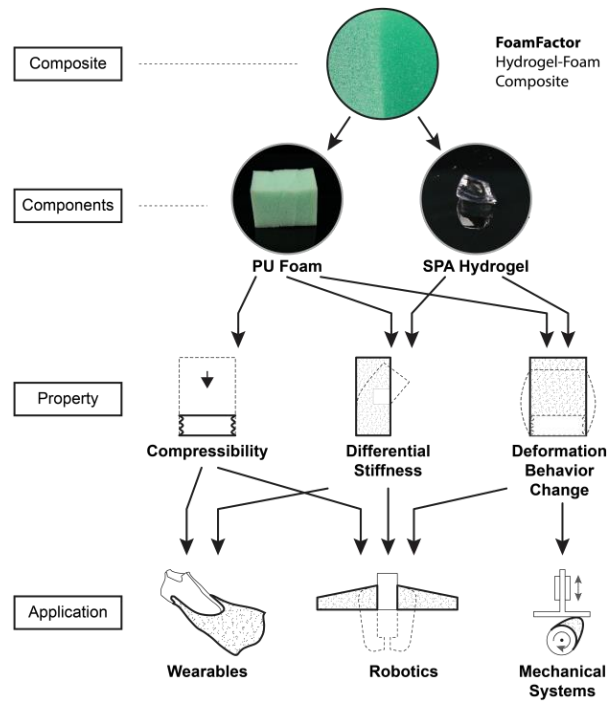

**Figure 3: The design space of FoamFactor for Interactive Systems.**

## MATERIAL PROPERTIES

The two components of our composite material - sodium polyacrylate (SPA) hydrogel and open-cell polyurethane (PU) foam - have contrasting structural properties. SPA is commonly used in commercial products due to its low cost and liquid absorption capability. SPA is condensed when dried. Upon hydration, it absorbs water rapidly within minutes and swells to hundreds of times (we measured 262 times) of its original volume. Due to this feature, SPA can be leveraged to create hydraulic actuators [39] or swelling-based shape changing materials [11]. Hydrated SPA is also strong against compression due to the incompressibility of water, but its polymer chains are weak against tension. On the other hand, open-cell PU foam is strong against tension but weak against compression due to its porosity.

Injecting SPA hydrogel into PU foam will create a composite structure like that of reinforced concretes. Each component withstands a specific type of load that they are strong against - hydrated SPA against compression and foam against tension. Additionally, the swollen hydrogel will fill the pores in the foam matrix, making it incompressible. In this state, the composite will buckle instead of getting compressed when subjected to a compressive load. When dehydrated, the properties of the foam will become dominant, making the composite compressible, soft, and conformable. Other than these mechanical properties, the ions in the hydrated SPA also made it conductive, capacitating sensing with it.

### Compressibility

The compressibility of FoamFactor samples (30 mm × 30 mm × 50 mm) were tested before and after hydration. The test samples were prepared and packaged in airtight bags and placed in a graduated cylinder filled with some water, then we exhausted the air with a vacuum pump (maximum vacuum: $3 \times 10^{-1}$ Pa) to compress the samples (Fig. 4). The water level changed in response to the volume loss of the samples. Divided by the original volume, we can get its compressibility - maximum volumetric change in percentage. The compressibility of FoamFactor samples differed largely before and after hydration. According to the results, the samples had an average volumetric compressibility of 79.3% before hydration and only 4.4% when hydrated. This is caused by the incompressible hydrated SPA filling the pores in the foam.

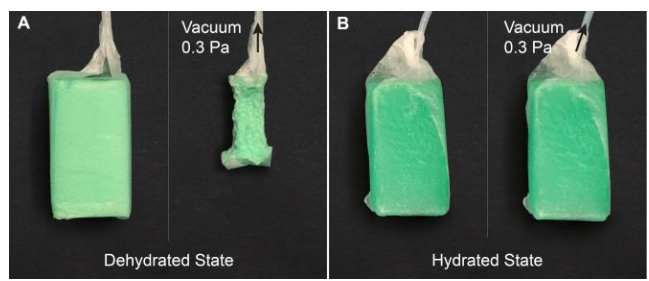

**Figure 4: Compressibility tests of a FoamFactor sample, relaxed (left) and compressed (right) in (A) dehydrated state and (B) hydrated state.**

### Stiffness Control

The pre-injection hydration ratio of the hydrogel will affect the stiffness of the final sample when it is fully hydrated. Figure 5A shows that the foams injected with more hydrated hydrogel had lower stiffnesses compared to those injected with less hydrated hydrogel. For later applications, we adopted 50% hydrated (pre-injection) hydrogel for fabrication as it provides the highest stiffness among the samples tested. Additionally, changing the infill ratio also caused the stiffness of the fabricated object to change.

The pre-injection hydration ratio of hydrogel was altered by supplying the hydrogel with a controlled amount of water. We first measured the maximum (100%) hydration ratio of the hydrogel by supplying it with excessive amount of water and left it to absorption for 24 hours, then measured its increase in weight. In our experiment, the maximum absorption was 212 grams of water for 1 gram of SPA. Based on this, we can produce 50%, 75% hydrated hydrogels by mixing 1 gram of SPA with 106 grams and 159 grams of water, respectively.

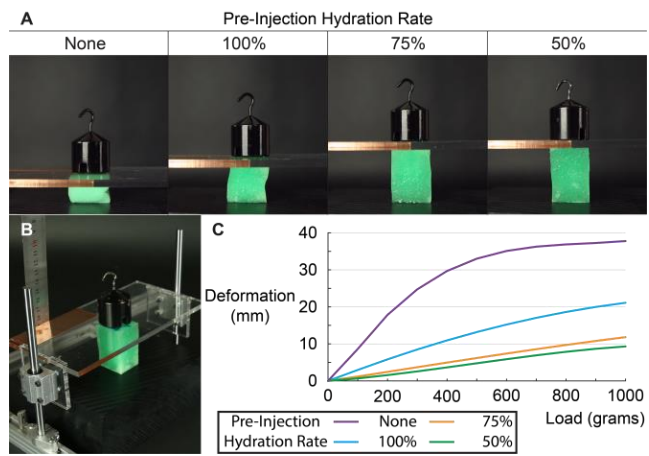

**Figure 5: Results from the stiffness tests. (A) Deformations of different samples (load: 200g). (B) Stiffness experiment platform. (C) Deformation-Load plot of the samples.**

A stiffness testing setup was designed for this experiment (Fig. 5B). Compressive loads were evenly applied onto the top surface of the samples (30 mm × 30 mm × 50 mm) through a flat plastic panel (platform), which could move freely and smoothly along two parallel sliding rails. The loads were applied incrementally, ranging from 100 grams to 1000 grams with a 100 grams increment. Figure 5C shows the measured deformations. Three samples injected with hydrogels of different pre-injection hydration ratios (50%, 75%, 100%) and a foam without injection were tested. Before the experiment, all injected samples were fully hydrated by soaking in water for 24 hours.

### Hydration Reversibility

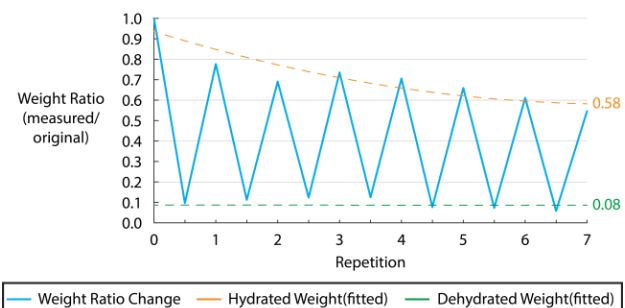

**Figure 6: Plot of repeated dehydrated and rehydrated weight compared to its original weight, showing the reversibility of salt water drying.**

We investigated salt-drying as to accelerate the dehydration of our material. Due to the water retaining ability of hydrogel and the bad convection inside the foams, drying a fully injected foam at room temperature in open air takes a long period of time. A FoamFactor piece of size 50 cubic mm took 120 hours to fully dehydrate. Even at an elevated temperature with ventilation, it may still take more than 72 hours to dehydrate. To save time in fabrication and iteration, we used table salt to expedite the drying process. During hydration, water moves from the surrounding into the hydrogel to balance the high concentration of sodium ion in SPA. We can leverage this phenomenon for fast drying during dehydration by increasing the sodium ion concentration in the surrounding to force water out of the hydrogel. Compared to water evaporation, this method may speed up the drying process by up to 50 times.

In this experiment, excessive salt was sprinkled over fully hydrated FoamFactor samples (50 cubic mm in size, 50% pre-injection hydration ratio hydrogel), which were then wrapped with a paper towel for 30 minutes, retrieved and squeezed to remove water and left in hot air (75 degree) for drying. Next, the samples were weighed and rehydrated for 30 minutes, and then weighed again to measure their weight loss. Figure 6 depicts the water retaining capacity of a foam after 7 iterations. For comparison, a foam of the same dimensions dried in hot air retained 87% of its fully hydrated weight. The micro-sized hydrogel granules are simply physisorbed inside the foam when dried, therefore it is inevitable that the very surface layer of the foam will lose contact of the hydrogel during rehydration in a water bath. The inner core of the foam, however, can trap the hydrogel in place after repeated dehydration/rehydration cycles, showing a less mass loss with increasing number of iterations. We conclude that adding salt to the foam will expedite the drying process as compared to air drying without sacrificing its water retaining capability.

**Conductivity and Strain Sensing**

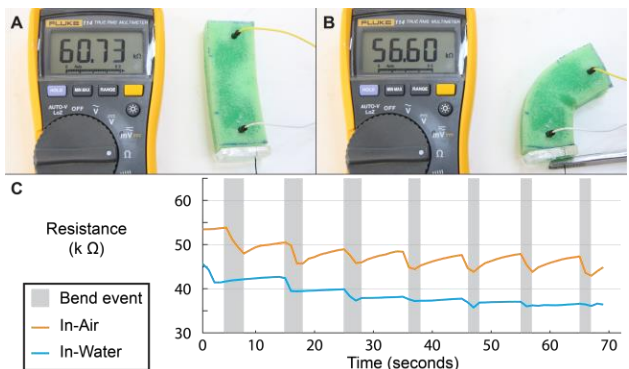

**Figure 7: Strain Sensing with FoamFactor. Bending events caused the resistance to drop. (A) Relaxed, resistance measured at 60.73 kΩ. (B) Bent, resistance measured at 56.60 kΩ. (C) Plot of resistance over time.**

SPA hydrogel is ionically conductive when hydrated and its resistance will change in response to deformations. Figure 7C shows the recorded values of the resistance disturbed by manual interferences both in-air and in-water. The resistance fell when the pieces were squeezed or bent. By tracking this drop, we can monitor interactions. This feature of the composite is valid when situated in both water and air. Note that the resistance will gradually increase to a certain point if left undisturbed.

## DESIGN AND FABRICATION WORKFLOW

### Hydrogel preparation

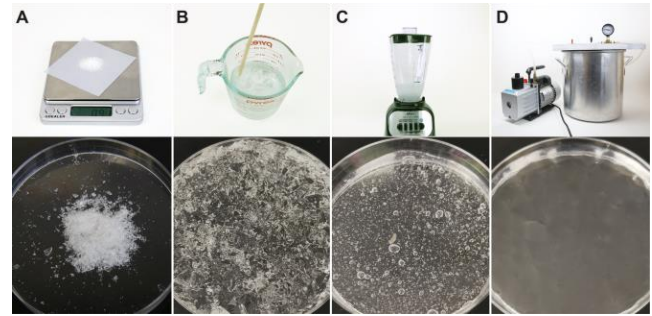

**Figure 8: Hydrogel preparation (top: equipment; bottom: processed hydrogel). (A) Weighing the dry SPA powder. (B) Mixing dry SPA with water. (C) Shredding granules with a food processor. (D) Degassing with a desiccator.**

In this paper we used off-the-shelf materials for fabrication and the granules of dry hydrogel were inconsistent in size (Fig. 8A). It is essential that the hydrogel has a comparably uniform granular size to ensure fabrication quality. Variations in the granular size and air bubble inclusions may cause an inconsistent injection rate. We tackled this by shredding and degassing the hydrogel before injection. A sample of hydrated hydrogel (Fig. 8B) was poured into a food processor and shredded until no chunks of hydrogel were observed (Fig. 8C). Shredded hydrogel was then degassed in a home-built desiccator (Fig. 8D). Bubble-free and homogeneous hydrogel can then be used for machine injection.

### Design Pipeline

We implemented our digital pipeline in Rhinoceros 3D with plugins (Grasshopper and Human UI). The pipeline as a script takes multiple solid bodies as input and generates G-code files for fabrication (Fig. 9). Prior to using this pipeline, users should first quantify the injection rate at different speed.

*Attribute Assignment*

Similar to CNC rasterizers, users must first assign the dimensions of a rectangular foam block as the matrix material, the target object can then be modeled directly in or imported into Rhino's CAD environment (Fig. 9A). Users can specify the hydration ratio for the current model and assign infill ratio to each body individually.

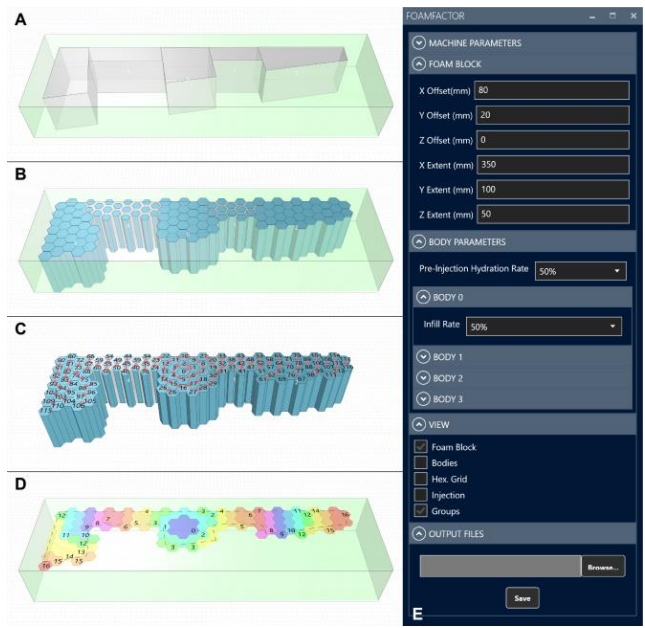

**Figure 9: Overview of the software interface. (A) Input shape. (B) User specifies the infill rate for each body. (C) Rasterization and toolpath planning. (D) Print job division. Each color corresponds to one print file. (E) Control panel for parametric design.**

*Hexagonal Rasterization*

The tool rasterizes the input shapes into multiple vertical hexagonal columns (Fig. 9B), in which the axes of the columns correspond to the injection movements of the needle tip, starting from the bottom and finishing at the top, moving at a constant speed. This pattern of tessellation gives the highest packing efficiency and approximates the hydrogel distribution in the foam.

*Toolpath Planning*

Upon contact, the foam may slightly compress before needle insertion, causing failure of penetration. The severity of this behavior differs between the location of insertion. At the center of the upper surface, the foam tends to compress less than it may have deformed if punctured around the edges. However, this problem can be alleviated if the insertion happens nearby previously injected columns, which act as a scaffolding preventing the foam to compress. Hence, the ideal injection order of the columns should start from the center of the surface and procedurally move outward to the edges (Fig. 10A). The pipeline orders the injections using the flood fill algorithm with the hexagon closest to the center of the grid as the root node (Fig. 9C).

A complete injection motion set (Fig. 10B) starts with horizontal movements at an offsetted height to position the needle, followed by a vertical movement of insertion. The needle will then move upward at a constant speed while dispersing the hydrogel. When the injection movement finishes, the needle will retrieve vertically to return to the offsetted height and proceed to the next injection.

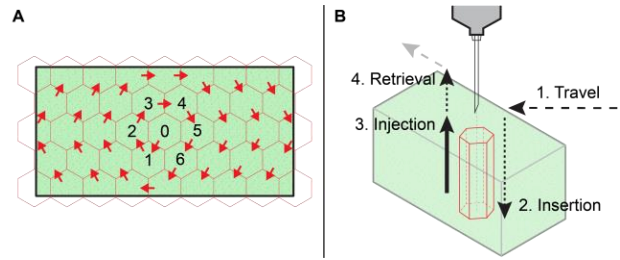

**Figure 10: (A) Printing tool path starting from the center. (B) A complete injection motion set.**

*Print Division and Compilation*

Due to the limited capacities of the pneumatic syringes, a print job may have to be divided into multiple files such that the injection volume for each file does not exceed the syringe capacity, and users can refill them in between files. Our interface visualizes this division and generates G-code files accordingly (Fig. 9D). The outputs also consist one G-code file for marking the outlines of the input bodies on the top face of the foam. The outlines are computed by projecting the bodies onto the top surface of the foam and can guide users to cut out the object with a hot wire cutter (Fig. 11E).

**Fabrication Platform**

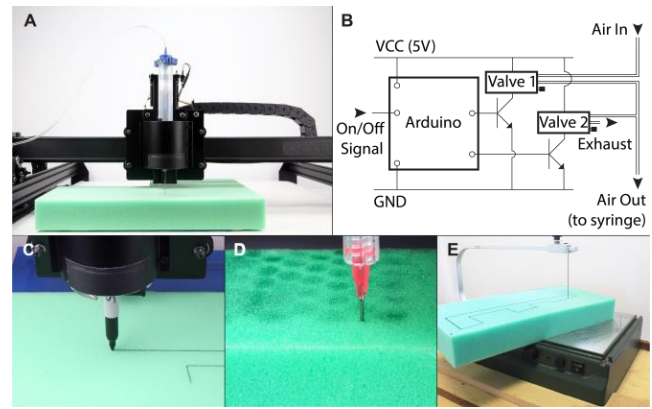

**Figure 11: The digital fabrication platform. (A) Gantry system with pneumatic injector. (B) Pneumatic controller circuit. (C) The machine marking an outline. (D) Injection. (E) Hot wire cutting.**

*Setup*

The digital fabrication process includes marking the outline, injecting hydrogel, and cutting away the excessive foam with a hot wire cutter. Our pipeline adopts a CNC gantry system with a customized pneumatic extrusion system for fabrication (Fig. 11A). The machine takes standard G-code as commands. The pneumatic system comprises a syringe, a needle, a controller, and a source of compressed air. The compressed air drives the piston in the syringe to push the material out. The gauge (18 ga.) of the needle was selected such that it is large enough to have a high injection rate while properly small to leave no holes on the foam when retrieved. The controller is an Arduino board that reads the machine's spindle signal to drive a pair of solenoid valves to alter the airflow (Fig. 11B). The air pressure (20 PSI) was set to give a consistent, high, and controllable injection rate. Lower pressure may cause the injection rate to be sensitive to changes in particle sizes, and higher pressure may cause the material to extrude too fast. Note that the needle gauge, the air pressure, and the injection movement speed together determine the injection rate and should be balanced to have a proper fabrication quality.

*Operation Parameters*

We favor machine parameters (e.g. traveling speed, insertion speed, and traveling height) that fabricate with haste and quality. The machine travels horizontally above the foam with a safe margin (5 mm) at a high speed (5000 mm/min) and inserts slowly (500 mm/min) to avoid damaging the foam. Additionally, the foam was taped to the platform to prevent displacements.

*Injection Rate Control*

The injection rate is not constant with different injection movement speeds due to the pressure built by the material accumulating at the needle tip. The injection rate also determines the hexagon size of the rasterization grid. This can be calculated with:

$$A = (Q \times t) / (S \times t) = Q / S$$

where S is the speed of the injection movement, Q is the injection rate, t is the time, and A is the hexagon area. Fig. 12 shows the relationship between different injection movement speeds and their corresponding hexagon sizes. Additionally, less hydrated hydrogel has a higher viscosity, resulting in a lower injection rate.

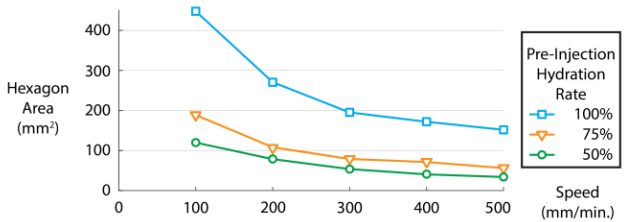

Figure 12: Plot of injection movement speed *S* versus hexagon area size A.

## APPLICATIONS

### Self-deployable Underwater Soft Gripper

Taking advantages of the stiffness-changing property of our composite material and its compressibility when dried, we designed an underwater gripper that has a compact packaging. The gripper arms made of FoamFactor were packaged with water dissolvable polyvinyl alcohol (PVA) films (Fig. 13A), such that the gripper arms were tightly packed when dried and could unpack and self-deploy in water (Fig. 13C). Figure 13D shows the gripper in motion underwater after deployment and hydration. The gripping motion was achieved by pulling the cables that ran through the arms and acted as tendons. In our implementation, the cables were pulled by a stepper motor-driven reel (Fig. 13B) and the actuations were controlled by an Arduino board.

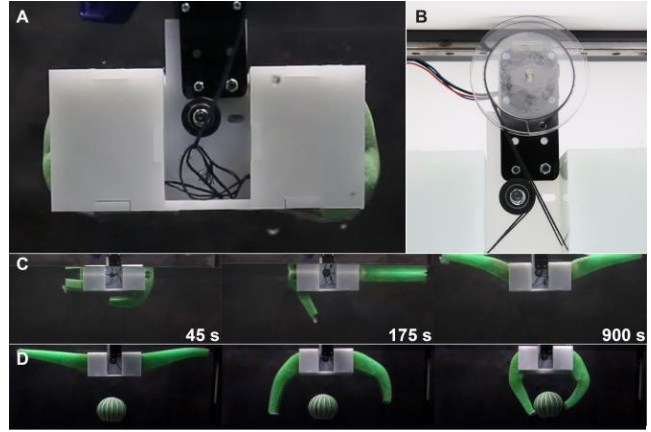

Figure 13: (A) Packaged gripper with arms compressed and stored. (B) Tendon control. (C) Gripper self-deployment process after being placed in water. (D) Gripping motion with hydrated arms.

The arms were differentially injected with hydrogel to create bending joints. As Figure 14A shows, the anisotropic distribution of hydrogel altered the bending behavior of the arm. Figure 14B provides a geometrical interpretation of the joint designs.

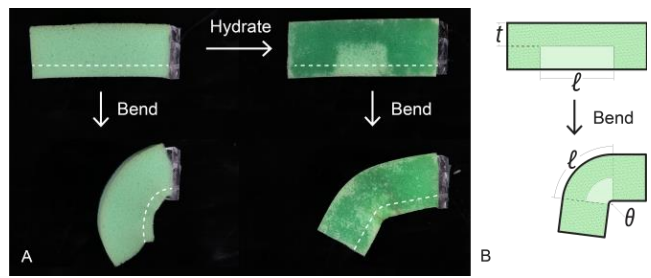

Figure 14: (A) Bending comparison of a dehydrated and hydrated Foamfactor arm. (B) Geometric interpretation of the bending angle and injection design. t determines the stiffness of the joint, ℓ determines the maximum bending angle θ.

### Amphibian Wheels

Properties of our composite material can be used in vehicular design, such as the wheels of amphibious cars. Figure 15 shows an amphibious car adopting FoamFactor as its wheels. On land, the pedals on the wheels were loosely packed for compactness while being conformable to the landscape. As soon as the car entered water, the PVA wrappings dissolved and released the pedals, which then were hydrated and became more rigid. We used additional foams without hydrogel injection to keep the car afloat on water (Fig. 15C and D). The wheels/pedals were driven by a stepper motor with an external controller. Figure 15E shows the transformation process of the wheels when placed in water.

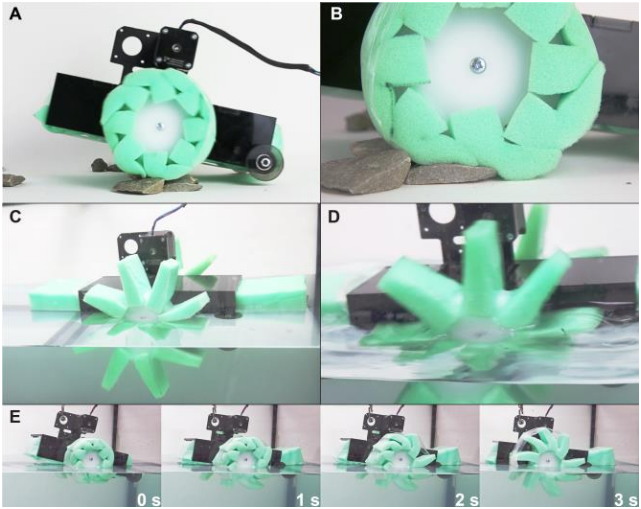

Figure 15: An amphibious car prototype. (A) On-land state of the car, with FoamFactor working as wheels. (B) The wheels conforming to the landscape. (C) In-water state of the car with FoamFactor working as pedals. (D) The pedals being more rigid and pushing water. (E) Transformation process of the wheels when placed in water.

**Contractible Swimming Fins**

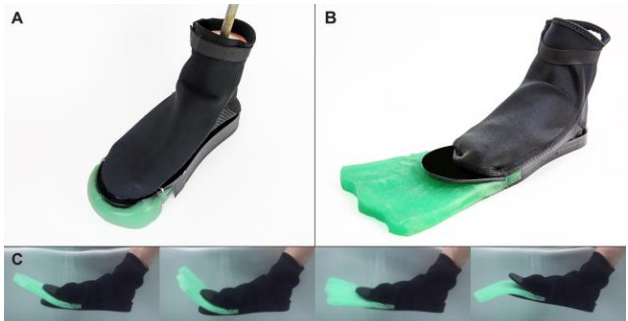

Figure 16: Convertible swimming fin. (A) Dehydrated (packed) state. (B) Hydrated (deployed) state. (C) Swimming.

We designed convertible wearables based on the properties of our material. Since the stiffness-changing is accompanied by hydration (i.e. submerged in water), we proposed to use the foam in amphibious augmentations, such as the contractible swimming fins (Fig. 16). On land, the fins made of FoamFactor were stored in the soles, covered and compressed by PVA films to make walking easy while wearing them. Upon contact with water, the films would dissolve and release the fins. The fins were then hydrated and stiffened to become functional.

**Motion Shifting Mechanical Systems**

Leveraging the changes in deformation behaviors of our material, we designed gearless transmission systems. By integrating FoamFactor into rigid mechanical systems, we achieved motion shifting without increasing their degrees of freedom. Here we demonstrate two instances of such integration. These components can potentially be adapted to design amphibious robots and gadgets.

*Linkage systems*

In linkage systems (Fig. 17), all rotary joints are passively driven by one active joint rotating at 360°. Instead of using hinges, the joints were replaced with FoamFactor, therefore soft and compliant when dried, and not interfering with rotations (Fig. 17C). When hydrated, it became stiffer and more rigid, prevented the joints from rotating, thus locked the system in place (Fig. 17D).

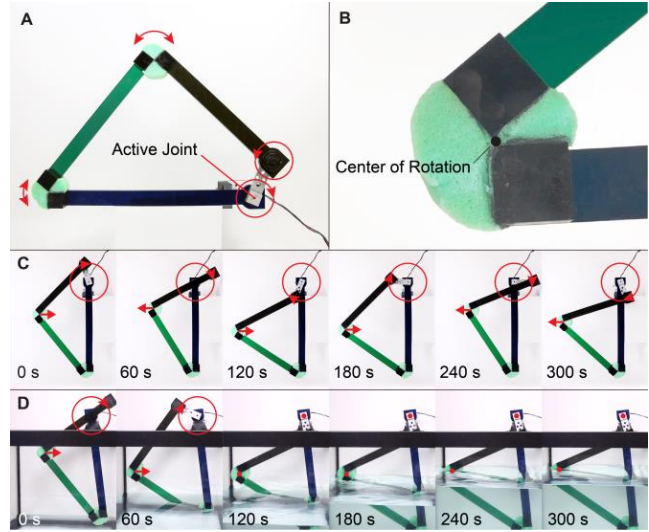

Figure 17: (A) Motions of a linkage system. (B) Detailed compliant joint design. (C) The linkage moved continuously when the compliant joints were dehydrated. (D) The hydrated joints stiffened and locked the linkage system.

*Cams*

A cam (Fig. 18) translates circular motions into linear motions. The plunger was subjected to a constant compressive load and was passively driven by the cam wheel's irregular radii. In our design, part of the cam wheel was replaced by FoamFactor, which would compress when it is dry, resulted in the plunger to maintain nearly motionless. When hydrated, the foam became incompressible and forced the plunger to move linearly in responses to its rotations.

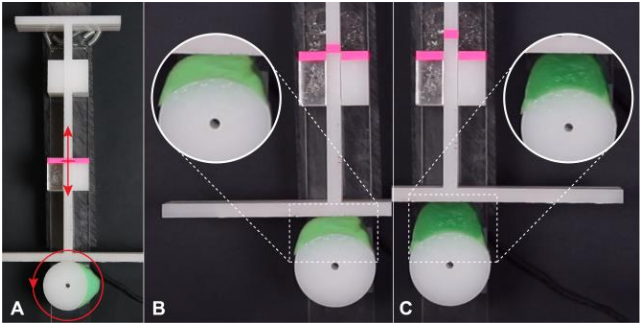

**Figure 18: (A)** Motions of a cam. **(B)** Dehydrated state. The foam was compressed by the plunger and the plunger moved upward slightly. **(C)** Hydrated State. The foam becomes stiffer and pushed the plunger upward more.

## DISCUSSION AND FUTURE WORK

### Variables of Hydrogel Dispersion

While all applications presented in this paper were fabricated with a single hydration ratio (50%), FoamFactor can be injected with hydrogels of different hydration ratio and may achieve more diverse functions. For instance, a beam injected with a mix of 50% and 100% hydration ratio hydrogel may exhibit a unique deformation behavior. A bending joint injected with 100% hydration ratio hydrogel may also be softer compared to one injected with 50% hydration ratio hydrogel.

### Geometrical Limitations

The geometric parameters of FoamFactor restrict the range of stiffness tunability. For instance, the fin we designed had a high surface area to endure thrust, but the flat outline reduced the stiffness when the fin was bent perpendicularly to the hydrodynamic movement. If we reduce the surface area, i.e. reduce the aspect ratio, we would have sacrificed the propulsion. The aspect ratio will therefore be an important factor to consider when designing a special functionality that involves stiffness tunability.

An artifact is vulnerable to dramatic deformations if possesses a high surface-area-to-volume ratio (e.g. being thin and flat, like the fin in Fig. 16). The hydrogel may be easily squeezed out of the foam during deployment, affecting the stiffness and compressibility. We thus need to control and minimize the surface-area-to-volume ratio without compromising its functionality. Furthermore, weight of the artifacts is another factor to consider in the design decision. Specifically, the amphibious car cannot stay neutrally buoyant without extra weight-bearing floatation due to the weight of the chassis and the motor.

### Functionalization of Hydrogel And/or Foams

The functions of FoamFactor depend on the properties of both hydrogel and foam. It will behoove us to incorporate more sophisticated material systems by either enriching the chemical complexity of the hydrogel or functionalizing the foam matrix to introduce higher level of modality. For instance, photoresponsive hydrogel, which may change the diffraction angle of light with corresponding volume changes, can be light sensitive and thus function as an optical sensor [3]. Furthermore, there are certain types of hydrogel that can endure salty environments [36], which will certainly broaden our applications to more natural environmental contexts (e.g. the ocean).

### Alternative Hydrogel Infusion Method

For all the samples and prototypes we have designed and made so far, hydrogel was injected into the foam in a linear fashion and hydrogel pillars inside the foam have the same shape and orientation. More Complex structures and distributions of hydrogel inside FoamFactor may be achieved with other hydrogel infusion methods. With different position, orientation and more structured distribution, the hydrogel inside the foam may form a more complicated 3D network.

Additionally, the PU foam has no chemical link to the hydrogel, causing mass loss during dehydration/rehydration cycles. It would greatly boost the repeatability of the rehydration/dehydration cycles of FoamFactor if we integrate chemical processes to material preparation, such as internally cross-linking the hydrogel to traverse the porous matrix or modifying the connection between foam and hydrogel.

## CONCLUSION

In this paper, a stiffness-changing and compressibility-tunable material was introduced and demonstrated, along with the workflow and digital-to-physical pipeline adapted to its fabrication. This material exhibits apparent changes in its stiffness and compressibility between hydration states, allowing for designing self-deployable and context-adaptive products. Although foam and hydrogel have been investigated under the context of soft actuations separately, to our knowledge, the composition method is simple yet novel. Our pipeline uses commercially available low-cost materials and machines, therefore is widely accessible to designers and researchers in different fields. We envision that this material will enrich the library of stiffness-changing interfaces, especially for contexts with hydration changes. Its potential applications include amphibious robots, convertible wearables, stiffness tunable components for transmission mechanisms and beyond.

## ACKNOWLEDGEMENTS

Insert texts here.

**REFERENCES**

[1] An, B., Tao, Y., Gu, J., Cheng, T., Chen, X. 'anthony', Zhang, X., Zhao, W., Do, Y., Takahashi, S., Wu, H.-Y., Zhang, T. and Yao, L. 2018. Thermorph: Democratizing 4D Printing of Self-Folding Materials and Interfaces. Proceedings of the 2018 CHI Conference on Human Factors in Computing Systems (New York, NY, USA, 2018), 260:1–260:12.

[2] Brown, E., Rodenberg, N., Amend, J., Mozeika, A., Steltz, E., Zakin, M.R., Lipson, H. and Jaeger, H.M. 2010. Universal robotic gripper based on the jamming of granular material. Proceedings of the National Academy of Sciences of the United States of America. 107, 44 (Nov. 2010), 18809–18814.

[3] Calvert, P. 2009. Hydrogels for Soft Machines. Advanced materials . 21, 7 (Feb. 2009), 743–756.

[4] Chenal, T.P., Case, J.C., Paik, J. and Kramer, R.K. 2014. Variable stiffness fabrics with embedded shape memory materials for wearable applications. 2014 IEEE/RSJ International Conference on Intelligent Robots and Systems (2014), 2827–2831.

[5] Cheng, N.G., Lobovsky, M.B., Keating, S.J., Setapen, A.M., Gero, K.I., Hosoi, A.E. and Iagnemma, K.D. 2012. Design and Analysis of a Robust, Low-cost, Highly Articulated manipulator enabled by jamming of granular media. 2012 IEEE International Conference on Robotics and Automation (May 2012), 4328–4333.

[6] Coelho, M., Ishii, H. and Maes, P. 2008. Surflex: A Programmable Surface for the Design of Tangible Interfaces. CHI '08 Extended Abstracts on Human Factors in Computing Systems (New York, NY, USA, 2008), 3429–3434.

[7] Elbaum, R., Zaltzman, L., Burgert, I. and Fratzl, P. 2007. The role of wheat awns in the seed dispersal unit. Science. 316, 5826 (May 2007), 884–886.

[8] Erb, R.M., Sander, J.S., Grisch, R. and Studart, A.R. 2013. Self-shaping composites with programmable bioinspired microstructures. Nature communications. 4, (2013), 1712.

[9] Evangelista, D., Hotton, S. and Dumais, J. 2011. The mechanics of explosive dispersal and self-burial in the seeds of the filaree, Erodium cicutarium (Geraniaceae). The Journal of experimental biology. 214, Pt 4 (Feb. 2011), 521–529.

[10] Follmer, S., Leithinger, D., Olwal, A., Cheng, N. and Ishii, H. 2012. Jamming User Interfaces: Programmable Particle Stiffness and Sensing for Malleable and Shape-changing Devices. Proceedings of the 25th Annual ACM Symposium on User Interface Software and Technology (New York, NY, USA, 2012), 519–528.

[11] Gladman, A.S., Matsumoto, E.A., Nuzzo, R.G., Mahadevan, L. and Lewis, J.A. 2016. Biomimetic 4D printing. Nature materials. 15, 4 (Apr. 2016), 413–418.

[12] Groeger, D., Chong Loo, E. and Steimle, J. 2016. HotFlex: Post-print Customization of 3D Prints Using Embedded State Change. Proceedings of the 2016 CHI Conference on Human Factors in Computing Systems (New York, NY, USA, 2016), 420–432.

[13] Heibeck, F., Tome, B., Della Silva, C. and Ishii, H. 2015. uniMorph: Fabricating Thin Film Composites for Shape-Changing Interfaces. Proceedings of the 28th Annual ACM Symposium on User Interface Software & Technology (Nov. 2015), 233–242.

[14] Ion, A., Frohnhofen, J., Wall, L. and Kovacs, R. 2016. Metamaterial mechanisms. Proceedings of the Conference of Army Physicians, Central Mediterranean Forces, held at the Institute [sic] superiore di sanita, Rome, 29th January to 3rd February 1945. Conference of Army Physicians, Central Mediterranean Forces (1945 ... (2016).

[15] Ion, A., Kovacs, R., Schneider, O.S., Lopes, P. and Baudisch, P. 2018. Metamaterial Textures. Proceedings of the 2018 CHI Conference on Human Factors in Computing Systems (New York, NY, USA, 2018), 336:1–336:12.

[16] Kan, V., Vargo, E., Machover, N., Ishii, H., Pan, S., Chen, W. and Kakehi, Y. 2016. Organic Primitives: Synthesis and Design of pH-Reactive Materials using Molecular I/O for Sensing, Actuation, and Interaction. arXiv [cs.HC].

[17] Katzschmann, R.K., Marchese, A.D. and Rus, D. 2016. Hydraulic Autonomous Soft Robotic Fish for 3D Swimming. Experimental Robotics: The 14th International Symposium on Experimental Robotics. M.A. Hsieh, O. Khatib, and V. Kumar, eds. Springer International Publishing. 405–420.

[18] Mac Murray, B.C., An, X., Robinson, S.S., van Meerbeek, I.M., O'Brien, K.W., Zhao, H. and Shepherd, R.F. 2015. Poroelastic Foams for Simple Fabrication of Complex Soft Robots. Advanced materials . 27, 41 (Nov. 2015), 6334–6340.

[19] Nakagaki, K., Vink, L., Counts, J., Windham, D., Leithinger, D., Follmer, S. and Ishii, H. 2016. Materiable: Rendering Dynamic Material Properties in Response to Direct Physical Touch with Shape Changing Interfaces. Proceedings of the 2016 CHI Conference on Human Factors in Computing Systems (New York, NY, USA, 2016), 2764–2772.

[20] Nakamaru, S., Nakayama, R., Niiyama, R. and Kakehi, Y. 2017. FoamSense: Design of Three Dimensional Soft Sensors with Porous Materials. Proceedings of the 30th Annual ACM Symposium on User Interface Software and Technology (New York, NY, USA, 2017), 437–447.

[21] Niiyama, R., Yao, L. and Ishii, H. 2013. Weight and Volume Changing Device with Liquid Metal Transfer. Proceedings of the 8th International Conference on Tangible, Embedded and Embodied Interaction (New York, NY, USA, 2013), 49–52.

[22] Olberding, S., Soto Ortega, S., Hildebrandt, K. and Steimle, J. 2015. Foldio: Digital Fabrication of Interactive and Shape-Changing Objects With Foldable Printed Electronics. Proceedings of the 28th Annual ACM Symposium on User Interface Software & Technology (New York, NY, USA, 2015), 223–232.

[23] Ou, J., Skouras, M., Vlavianos, N., Heibeck, F., Cheng, C.-Y., Peters, J. and Ishii, H. 2016. aeroMorph - Heat-sealing Inflatable Shape-change Materials for Interaction Design. Proceedings of the 29th Annual Symposium on User Interface Software and Technology (Oct. 2016), 121–132.

[24] Ou, J., Yao, L., Tauber, D., Steimle, J., Niiyama, R. and Ishii, H. 2013. jamSheets: Thin Interfaces with Tunable Stiffness Enabled by Layer Jamming. Proceedings of the 8th International Conference on Tangible, Embedded and Embodied Interaction (New York, NY, USA, 2013), 65–72.

[25] Qi, J. and Buechley, L. 2010. Electronic Popables: Exploring Paper-based Computing Through an Interactive Pop-up Book. Proceedings of the Fourth International Conference on Tangible, Embedded, and Embodied Interaction (New York, NY, USA, 2010), 121–128.

[26] Reyssat, E. and Mahadevan, L. 2009. Hygromorphs: from pine cones to biomimetic bilayers. Journal of the Royal Society, Interface / the Royal Society. 6, 39 (Oct. 2009), 951–957.

[27] Robertson, M.A. and Paik, J. 2017. New soft robots really suck: Vacuum-powered systems empower diverse capabilities. Science Robotics. 2, 9 (2017), eaan6357.

[28] Sahoo, D.R., Neate, T., Tokuda, Y., Pearson, J., Robinson, S., Subramanian, S. and Jones, M. 2018. Tangible Drops: A Visio-Tactile Display Using Actuated Liquid-Metal Droplets. Proceedings of the 2018 CHI Conference on Human Factors in Computing Systems (New York, NY, USA, 2018), 177:1–177:14.

[29] Sareen, H., Umapathi, U., Shin, P., Kakehi, Y., Ou, J., Ishii, H. and Maes, P. 2017. Printflatables: Printing Human-Scale, Functional and Dynamic Inflatable Objects. Proceedings of the 2017 CHI Conference on Human Factors in Computing Systems (New York, NY, USA, 2017), 3669–3680.

[30] Simon, T.M., Smith, R.T. and Thomas, B.H. 2014. Wearable Jamming Mitten for Virtual Environment Haptics. Proceedings of the 2014 ACM International Symposium on Wearable Computers (New York, NY, USA, 2014), 67–70.

[31] Simon, T.M., Thomas, B.H. and Smith, R.T. 2015. Controlling Stiffness with Jamming for Wearable Haptics. Proceedings of the 2015 ACM International Symposium on Wearable Computers (New York, NY, USA, 2015), 45–46.

[32] Tadakuma, K., Arimie, E.F.M. and Tadakuma, R. 2013. Hyper flexible robot with variable stiffness and shape. 2013 IEEE/ASME International Conference on Advanced Intelligent Mechatronics (Jul. 2013), 1318–1323.

[33] Van Meerbeek, I.M., Mac Murray, B.C., Kim, J.W., Robinson, S.S., Zou, P.X., Silberstein, M.N. and Shepherd, R.F. 2016. Morphing Metal and Elastomer Bicontinuous Foams for Reversible Stiffness, Shape Memory, and Self-Healing Soft Machines. Advanced materials . 28, 14 (2016), 2801–2806.

[34] Wang, G., Cheng, T., Do, Y., Yang, H., Tao, Y., Gu, J., An, B. and Yao, L. 2018. Printed Paper Actuator: A Low-cost Reversible Actuation and Sensing Method for Shape Changing Interfaces. Proceedings of the 2018 CHI Conference on Human Factors in Computing Systems (Apr. 2018), 569.

[35] Wang, W., Yao, L., Zhang, T., Cheng, C.-Y., Levine, D. and Ishii, H. 2017. Transformative Appetite: Shape-Changing Food Transforms from 2D to 3D by Water Interaction Through Cooking. Proceedings of the 2017 CHI Conference on Human Factors in

Computing Systems (New York, NY, USA, 2017), 6123–6132.

[36] Xu, J., Ren, X. and Gao, G. 2018. Salt-inactive hydrophobic association hydrogels with fatigue resistant and self-healing properties. Polymer. 150, (2018), 194–203.

[37] Yao, L., Niiyama, R., Ou, J., Follmer, S., Della Silva, C. and Ishii, H. 2013. PneUI: Pneumatically Actuated Soft Composite Materials for Shape Changing Interfaces. Proceedings of the 26th Annual ACM Symposium on User Interface Software and Technology (New York, NY, USA, 2013), 13–22.

[38] Yao, L., Ou, J., Cheng, C.-Y., Steiner, H., Wang, W., Wang, G. and Ishii, H. 2015. bioLogic: Natto Cells as Nanoactuators for Shape Changing Interfaces. Proceedings of the 33rd Annual ACM Conference on Human Factors in Computing Systems (Apr. 2015), 1–10.

[39] Yuk, H., Lin, S., Ma, C., Takaffoli, M., Fang, N.X. and Zhao, X. 2017. Hydraulic hydrogel actuators and robots optically and sonically camouflaged in water. Nature communications. 8, (Feb. 2017), 14230.

[40] Zubrycki, I. and Granosik, G. 2017. Novel Haptic Device Using Jamming Principle for Providing Kinaesthetic Feedback in Glove-Based Control Interface. Journal of Intelligent and Robotic Systems. 85, 3 (Mar. 2017), 413–429.